\documentstyle[a4wide,epsfig,12pt]{article}

\begin{document}     
\parindent 0pt
\parskip 5pt

\vskip 10pt
\noindent
\centerline{\large \bf TeV $\gamma$-ray Observations of the Crab and Mkn 501}
\vskip 4pt
\centerline{\large \bf during Moonshine and Twilight.
\footnote{submitted to: {\bf Astroparticle Physics}}}
\par\vspace{0.5cm}

D. Kranich$^1$,
R. Mirzoyan$^1$,
D. Petry$^{1,8,10}$,
B.C. Raubenheimer$^2$,
F. Aharonian$^3$,
A.G. Akhperjanian$^4$,
J.A.Barrio$^{1,5}$,
C. Beck$^1$,
K. Bernl\"ohr$^{3,9}$,
H. Bojahr$^8$,
J.L. Contreras$^5$,
J. Cortina$^{1,5}$,
A. Daum$^3$,
T. Deckers$^7$,
S. Denninghoff$^1$,
V. Fonseca$^5$,
J. Gebauer$^1$,
J.C. Gonzalez$^5$,
G. Heinzelmann$^6$,
M. Hemberger$^3$,
G. Hermann$^3$,
M. Hess$^3$,
A. Heusler$^3$,
W. Hofmann$^3$,
H. Hohl$^8$,
D. Horns$^6$,
A. Ibarra$^5$,
R. Kankanyan$^3$,
M. Kestel$^1$,
O. Kirstein$^7$,
C. K\"ohler$^3$,
A. Konopelko$^3$,
H. Kornmayer$^1$,
H. Krawczynski$^{3,6}$,
H. Lampeitl$^3$,
A. Lindner$^6$,
E. Lorenz$^1$,
N. Magnussen$^8$,
H. Meyer$^8$,
A. Moralejo$^5$,
L. Padilla$^5$,
M. Panter$^3$,
R. Plaga$^1$,
A. Plyasheshnikov$^3$,
J. Prahl$^6$,
G. P\"uhlhofer$^3$,
G. Rauterberg$^7$,
C. Renault$^3$,
W. Rhode$^8$,
A. Roehring$^6$,
V. Sahakian$^4$,
M. Samorski$^7$,
D. Schmele$^6$,
F. Schr\"oder$^8$,
W. Stamm$^7$,
H. V\"olk$^3$,
B. Wiebel-Sooth$^8$,
C. Wiedner$^3$,
M. Willmer$^7$,
H. Wirth$^3$,
W. Wittek$^1$\\

{\it
$^1$ Max-Planck-Institut f\"ur Physik, F\"ohringer Ring 6,
     D-80805 M\"unchen, Germany \\
$^2$ Space Research Unit, PU for CHE, 2520 Potchefstroom, South Africa\\
$^3$ Max-Planck-Institut f\"ur Kernphysik, Postfach 103980,
     D-69029 Heidelberg, Germany \\
$^4$ Yerevan Physics Institute, Alikhanian Br. 2, 375036 Yerevan, Armenia \\
$^5$ Universidad Complutense, Facultad de Ciencias F\'{\i}sicas,
     Ciudad Universitaria, E-28040 Madrid, Spain \\
$^6$ Universit\"at Hamburg, II. Institut f\"ur Experimentalphysik,
     Luruper Chaussee 149, D-22761 Hamburg, Germany \\
$^7$ Universit\"at Kiel, Institut f\"ur Kernphysik, Leibnitzstr. 15,
     D-24118 Kiel, Germany \\
$^8$ Universit\"at Wuppertal, Fachbereich Physik, Gau{\ss}str.20,
     D-42097 Wuppertal, Germany \\
$^9$ Now at Forschungszentrum Karlsruhe, P.O. Box 3640, D-76021 Karlsruhe \\
$^{10}$ Now at Universidad Aut\'{o}noma de Barcelona, Institut de
        F\'{\i}sica d'Altes Energies, E-08193 Bellaterra, Spain
}

\begin{abstract}
TeV $\gamma$-ray signals from the Crab Nebula and Mkn 501 were detected with 
the
HEGRA CT1 imaging Cerenkov telescope during periods when the moon was
shining and during twilight. This was accomplished by lowering the high 
voltage 
supply of the
photomutipliers in fixed steps up to 13\%.
No other adjustments were made and no filters were used. Laser
runs could not establish any non-linearity in the gain  of the
individual pixels, and the trigger rate was uniform over the whole
camera. The energy threshold was increased by up to a factor of two, 
depending on the amount of HV reduction.
In a series of observations lasting
11.7 hours, a signal with a 3.4$\sigma$ significance was detected from the 
Crab. 
During the 1997 multiple flare episode of Mkn 501 a 26$\sigma$ combined 
excess
has been recorded during 134 hours of observations under various 
moonshine/twilight conditions.
The results show that this
technique can easily be adapted to increase the exposure of a source, which 
is important for sources showing rapid time variability
such as AGNs or GRBs.
Observations can be made up to $\sim20^\circ$ 
angular separation from the moon and
until the moon is 85\% illuminated (ten to eleven days before and after new
moon), as well as during 20 to 40 minutes during twilight, before the 
commencement of 
astronomical darkness. 


\end{abstract}

\section{Introduction}

Any ACT telescope operating under the strict condition of no moonlight
during observations, can reach a theoretical duty cycle of 18\% per year
(Dawson and Smith, 1996) at a latitude of 40$^\circ$. 
If this strict criterion
is relaxed to observations under partial moonlight (e.g.~70\% of the
moon illuminated, i.e. a period of nine days before or after new moon),
an increase of the duty cycle to 24\% is possible. A further increase 
is possible when observations are extended into twilight time, 
when similar conditions exist.
In view of recent reports of source variability, it
is important to monitor a source as long as possible.

Furthermore, with the advent of the next generation of very large, low
threshold energy, imaging telescopes, any increase of the  duty cycle will make
the financial investment and scientific yield more attractive. The method 
should also have a rapid reaction time ($\sim 1$ minute) to  enable the 
observation of GRBs (Gamma Ray Bursts). 
It is
therefore necessary to investigate ways to increase the duty cycle of
ACT telescopes in such a way that it can be easily realized without too
great a loss in sensitivity.

Earlier, Pare et al. (1991) reported the use of solar blind 
photomultipliers (PMTs, with
sensitivity limited to the UV range) to observe during moonshine. The
approach proved successful but the threshold energy was increased by a
factor of 3.5 with respect to measurements with normal PMTs during
no-moon conditions. Although such an approach is possible, it is
expensive and time consuming since a second camera must be available and
should be interchanged with the normal one to operate during moonshine.

A second approach is to use a UV sensitive filter in front of normal
PMTs to block out most of the scattered light from the moon. Such a
system was successfully  tested by  Chantell et al. (1995). Operations
were extended up to full moon and limited to positions more than 10$^\circ$
from the moon. Successful detection of the Crab was possible after
additional software cuts (apart from the normal supercuts). The energy
threshold was 3.5 times the no-moon value and  10 times more
observation time was needed to reach a specific significance. It is
clear  that a more modest approach is needed to limit the increase of
threshold and time. Bradbury et al. (1996) suggest the use of wavelength
shifters which could increase the UV sensitivity and could be used alone
or in combination with the abovementioned filter system, partly
counterbalancing the increases.

It is known from measurements in the U-band (300 to 400 nm), which
represents the most sensitive wavelength band of ACT (Atmospheric Cerenkov 
Technique)  telescopes, that
the NSB (night sky brightness) increases by a factor of three to five
(see Figure 1) during a
half illuminated moon (compared to no-moon conditions), rising
to a factor of 30 - 50 during  full moon (Dawson and Smith, 1996; Schaefer, 
1998). This increase depends on various factors including the telescope 
altitude, moon angle, 
zenith angle and 
atmospheric composition as well as the aerosol content.
As most
imaging telescopes are operating on a double trigger threshold (i.e. the
hardware trigger which is determined by the fluctuations in the NSB, and
a much higher image threshold (e.g. $\sim30$ photo-electrons for the HEGRA 
CT1 telescope)), we
expect  that increases in the NSB of up to a factor of 10 will not have a
marked effect  on the quality of the produced images. A slight increase
of the hardware threshold (e.g. through lowering of the PMT high voltage (HV))
 due to
moonlight does not imply significant changes of
 the image threshold and thus we expect
no major changes in data analysis to occur when we observe during
moonshine.

In an attempt to lower the energy threshold, Quinn et al. (1996)
increased the HV of the PMTs by 40\% and the produced images in the camera were
similar to those obtained under normal conditions. This is an indication
that the normal supercuts analysis (to obtain an enhanced $\gamma$-ray signal,
 see Petry et al. (1996))
is robust to changes in the system gain, provided that the pixel
response is uniform. The image parameters do however change when an UV
filter system is used (Chantell et al., 1995). This is due to a change
in the spectral composition of the Cerenkov light.

The above discussion shows that although various techniques were
investigated, an effective and simple method to increase observation
time is still lacking.  In the following we report on such a technique
and illustrate it with observations of the Crab Nebula 
and Mkn 501 with the HEGRA CT1 telescope.

\section{Observations}
\subsection{Exploratory measurements}

Since the differential spectrum of the NSB, sun and moonlight peak in the 
yellow to red region of the spectrum (e.g. 
Dobber, 1998), 
the bulk of this light is not registered by the blue sensitive PMTs used in 
the ACT. Furthermore, most imaging telescopes are equipped with Winston cones 
on their cameras, preventing most of the scattered light from atmospheric 
particles (both Rayleigh and Mie scattering) and the environment to enter 
the detection system. Scattered light from high altitude haze and ice crystals
 are also 
excluded since no observations are conducted under these conditions since it 
make the data unreliable (shown e.g., by Snabre, et al., 1998). It is 
therefore believed that most imaging ACT telescopes may regularly operate 
during 
these conditions of increased illumination without deterioration of their 
detectors, provided that they operate with low to medium gain PMTs. 
By using an atmospheric extinction program (Schaefer, 1998) which include 
Rayleigh and Mie scattering, it is clear from Figure 1, that apart from an 
exclusion zone 
around the moon (varying from $\sim20^\circ$ to $\sim40^\circ$, depending 
on the illumination of the moon and the haziness of the sky), a relative 
constant NSB-level, as a 
function of zenith angle, may be expected. 

All measurements were conducted with the HEGRA CT1 telescope with
its 5 m$^2$ reflector and 127 pixel camera operating at a threshold
energy of 1.7 TeV during December 1996 (Mirzoyan et al. (1994) and Rauterberg et al. (1995)).
The usual hardware trigger of at least 2 tubes triggering at 15
photo-electrons was used with the low gain PMTs. 
The 10 stage EMI 9083 A PMTs are operated with only 8
stages and  AC coupled
fast amplifiers compensate for the reduced gain. This operational mode
allows one to circumvent large and damaging anode currents in the PMTs from 
e.g. the NSB, as well as scattered moon- and sunlight (during dusk and dawn). 
The 
measurements were conducted during December 1996 whilst the moon was  nine
days old ($-12.5$ visual magnitude, 70\% illumination). This implied an 
increase in the NSB by a factor of 20 (see Figure 1). The telescope was
pointed in a direction $\sim90^\circ$
 away from the moon. 
Without any adjustment
to the PMTs, the accidental trigger rate (ATR) 
and the average PMT current increased as expected.

The HV of the PMTs, and thus the system gain was lowered to ensure
minimal PMT fatigue and to lower the
accidental trigger rate (ATR). The small signals
causing the increased ATR disappeared rapidly and at a HV reduction of 4\%, 
the ATR was down to a manageable
0.09 Hz, and the  average PMT current was 8 $\mu$A. The third magnitude star
$\zeta$ Tauri was clearly visible in the camera and cosmic ray events could
easily be recognised against a low background. The result of laser
calibration runs assured us that the individual pixels were responding
linearly. This was borne out by off-line analysis which shows uniform
triggering throughout the camera. The raw trigger rate was 65 - 70\% of that
under no-moon conditions, indicating an estimated increase in the
energy threshold from 1.7 to 2.4 TeV.
From the gain characteristics of our PMTs ($\Delta g/\Delta U = 2/140$V
at the operational voltage of 1080 V, averaged over all PMTs), we calculate 
a gain reduction of 25\% for
a HV reduction of 4\%. This, in turn resulted in a 40\% flux reduction for a 
power law
spectrum with $\alpha = - 1.6$, which is in good agreement with the 
observed value of 38\%.
It should be noted that neither is the PMT current exactly proportional to the 
NSB photon flux (due to non-linear gain effects and base-line shifts caused by 
AC coupling), nor is the trigger rate directly proportional with the PMT gain. 

With the telescope apparently operating normally, it was pointed
 gradually  closer to the moon. No significant changes in the ATR or PMT 
currents
could be detected up to 30$^\circ$ from the moon, where the NSB started 
to increase rapidly, in accordance with Figure 1. This preliminary 
measurements 
indicated that the
telescope could be operated over a large region of the sky without any
marked influence from the moon.

Further measurements showed that the abovementioned operating conditions
can be maintained until the moon is 85\% illuminated (11 days before and
after new moon), provided that the moon is not approached closer than
20$^\circ$. From Figure 1, it is therefore clear that the camera can handle 
NSB increases of up to a factor of $\sim$50.

\subsection{Crab observations}

Under the operating  conditions described above, we observed the Crab
Nebula (the only ACT source with a constant flux) 
during five nights under varying 
moon conditions (from 5 to 9
days old and approaching the moon itself up to 22$^\circ$). Data at zenith
angles smaller than 30$^\circ$
 were used (see Table 1 for further detail).

A total of 11.7 hours of observations on the Crab was analysed.
Applying normal supercuts (Petry et al., 1996) resulted in a
significance S (in standard deviations) of 0.8 $\sqrt{t}$ 
compared to 1.3 $\sqrt{t}$
(with $t$ in hours) under no-moon conditions. No large increase in
observation time is therefore needed to reach the same significance as
under no-moon conditions. 
The fact that our
significance, S, is smaller under moon conditions, indicates that there is
indeed a change in the parameters for $\gamma$-ray selection, indicating the 
need of  
software optimisation of the analysis of moonshine (MS) data. 

In Figure 2, the ALPHA-distribution is shown. 
It is clear that it is
similar (but flatter - See Section 2.3) than that observed from other sources 
under normal dark moon
conditions. To determine the $\gamma$-ray excess, we used 80 h of background
data compiled over the previous two years by CT1 (Petry et al., 1997).
The justification for this procedure stems from the excellent
agreement of the ON- and OFF-data for ALPHA $> 20^\circ$. 
 It is
however essential that simultaneous background measurements be made to
establish any effect of the increased NSB on the background (See Section 2.3).
To check for these possible biases, the number of
expected background events was determined, in a second approach, using the
region $20^\circ < ALPHA < 80^\circ$ of the 
ON-data instead of OFF-data (expected
background events: N = {N$_{on}$} ($20^\circ < ALPHA < 80^\circ$) / 6). 
With this method a slightly more significant result was obtained 
(see Table 1). 

The resulting $\gamma$-ray rate of 4 h$^{-1}$ is 62\% of the 6.4 h$^{-1}$ 
seen by the HEGRA CT1
telescope during the same period from the Crab under normal, no-moon 
conditions 
at a threshold of 1.7 TeV.  
This  should be compared with only 7\% of the no-moon rate reported by Chantell
et al. (1995) with their filter system. Thus, our technique provides a
smaller increase in threshold and can easily be adapted by any imaging
telescope to increase the exposure of sources.

\subsection{Mkn 501 observations}

From March to September 1997 the AGN Mkn 501 showed  strong, variable
$\gamma$-ray emission at an average flux twice that of the Crab 
(see e.g., Kranich, et al. (1997) and Protheroe, et al. 
(1998)). Having proven that the moonshine technique is working, this event 
provided the possibility of increasing our exposure of the source and, at the 
same time, investigate the moonshine observations more extensively.
At the end of April 1997, the nominal operating HV of the PMTs was increased 
by 6\% in order to 
re-establish the sensitivity to that of 1994. This adjustment was needed due 
to normal ageing of the PMTs, with a 
gain reduction of 20-30\% after two years of operation. 

With the experience of the Crab observations, we adopted a more conservative 
and refined observing strategy for the moonshine observations: The PMTs were
running
at nominal voltage up to a 20\% illuminated  moon (3 nights), a 6\% reduction 
in HV up 
to 70\% illumination (5 nights), an
9\% reduction in HV up to 90\% illumination (2 nights) and a 13\% reduction 
up to 95\% illumination (one night). This scaling of HV reduction correlates 
with the increase of NSB 
during increasing moon luminosity, and exclude only the night during full moon.
This strategy is quantified in Table 2, together with the expected threshold 
energies of the various HV settings.
A limit on the average PMT current of 12 $\mu$A (20 $\mu$A 
maximum)
was set. This implied that a further HV reduction was made as soon as this 
limit was reached - this occurred a number of times during the same night (e.g. 
MJD 50615 and 50641 in
 Figure 4)
 as the source was approaching the moon. 

With this observing strategy, we increased the total exposure of the source by
 56\% compared to the normal dark moon observations. The final data set 
consists of 28 nights of MS 
data, 41 with both MS and dark moon data (making
 a comparison possible) and 52 nights with only dark moon data. Due to this 
additional observations, the Mkn 501 data set of the HEGRA CT1 telescope is 
the most complete of all ACT observations of this event (see Table 2 
for detail). 

The most important characteristics of the observations are summed up in 
Table 2.
From this Table the following conclusions may be drawn: 
(i) The quality factor, 
Q, describing the $\gamma$/hadron separation capability,

\centerline {Q = $[N_{on}/T]/{\sqrt{N_{b}/T}}$}

(with $N_{on}$ and $N_{b}$ respectively the ON-source and OFF-source events 
with ALPHA $< 10^{\circ}$)  decreases with decreasing HV. 
This is understandable 
since the recorded images will become smaller with 
increasing HV reduction, due to higher tail cuts (which is a result of the 
increase of PMT noise caused by moonlight). 
Furthermore, the moonshine will produce additional noise which cannot be
filtered out with the normal supercuts analysis.   
(ii) Comparing the two sets of HV settings which contain both dark moon and MS 
data, it is clear that Q is the same, assuring us that the additional light 
due 
to moonshine does not have a marked influence on the $\gamma$-hadron 
separation. 
(iii) The Q-value of the Crab observations fits in with the general 
dependency of Q on the HV reduction.
(iv) The background rate, at nominal HV, was 20\% higher with moonlight than 
without. This increased to 55\% at a 6\% HV reduction. It can therefore be 
inferred 
that this rate will continue to increase with higher HV reductions, 
contributing more and more to the background and possibly dilutes the 
signal. Care 
should therefore be taken with HV reductions larger than 10\% and Monte Carlo 
studies are indicated to investigate the effect of moonlight on the normal 
supercuts analysis.

In Figure 3, the ALPHA-distribution for 244 h dark moon
observation and 134 h moonshine observations is shown, 
divided into six subsets, 
according to the applicable HV setting. To maximise the source exposure, we 
used as OFF-source  
data a sample taken during dark nights, as discussed in Section 2.2. 
We also recorded a
small sample of OFF-data during moonshine which was in good
agreement with the ALPHA-distribution of the dark night OFF-source data. 
Comparing the various panels, we conclude the following:
(i) The background-region (ALPHA $>20^{\circ}$) has the same shape 
for all the 
cases, independent of HV reduction or the presence of moonlight.
(ii) Comparing Figures 3(a) with 3(b) as well as 3(c) with 3(d), an increase 
in the background rate (i.e. ALPHA $>20^{\circ}$) is evident as soon as 
moonlight contributes to the NSB. 
 (iii) It is clear from all six panels that the shape of the ALPHA-excess does 
became flatter with decreasing HV.
This is  
attributable to additional noise in all the camera pixels due to moonlight. 
A confirmation of this effect is apparent when Figure 2 is compared with 
Figure 3(f): In both cases the flattening of the ALPHA-excess is similar, 
illustrating that this flattening is mainly determined by the addition of 
moonlight to the detector and is not due to the fact that the Crab has a 
lower flux than Mkn 501.

We therefore conclude that the supercuts analysis is robust enough for 
observing during moonshine in the way described. Care should 
however be taken when the HV is reduced by more than 10\%, due to low Q-values.
This excludes observations during the last two to three nights before full 
moon.  

Figure 4 shows the light curve for Mkn 501, including the MS/twilight data, 
up to zenith angles of $60^{\circ}$. The fluxes should be considered as 
preliminary due to a shortage of Monte Carlo data at large zenith angles. The
errors for the MS data are generally larger because 
(a) the MS measurements were mostly of shorter duration, and (b)
for a coherent presentation the integral flux data were calculated for 1.5 
TeV, i.e. the MS data were extrapolated, using a power law coefficient of
$-1.5$, assuming the threshold  energies in Table 2.
 During the 216 days of the multiple flare, we were able to collect data
during all nights with clear weather, excluding only the nights during full 
moon. This is a good example of the value of MS observations. 

During 39 nights we recorded both dark moon and MS data and a
comparison between the fluxes for the resultant 57 pairs of moon/no-moon data 
can be made. Although Mkn 501 
showed 
flux variability on a few hours time scale, the procedure of selecting pairs 
from the same night minimise this effect. From Figure 5 a good agreement is 
apparent with a correlation coefficient of 0.81. From the fitted line a 
gradient of 0.63 was calculated. This value increases to 0.83 when only data 
at the nominal HV is analysed. These two values should be compared with a 
theoretical 
value of 1. 
The fact that the MS data show larger fluxes than the corresponding no-moon 
data 
may be attributable to an overestimation of the effect of HV reduction on the 
energy threshold, the shortage of Monte Carlo simulations at large zenith 
angles as well as the use of non-optimised, normal supercuts.
This effect is under further investigation.

\subsection{Observations during Twilight}

With the successful implementation of MS measurements, it was realized that 
additional observations could be possible during twilight. The conventional 
practice of ACT telescopes is to start observing after Astronomical Twilight 
(when the sun is more than $18^\circ$ below the horizon). Again, applying the 
atmospheric model of Schaefer (1998), it became clear that observations 
could indeed be extended into twilight time. Depending on the pointing 
direction of the telescope, the observations could start as soon as  
Nautical Twilight (the sun reaching $12^\circ$ below the horizon). 
This could add 20 to 40 minutes of additional observations, depending on the 
latitude of the observatory. 

It was decided to make single runs, lasting 20 
minutes, just before Astronomical Twilight, at the nominal HV of the PMTs.  
Ten such observations were made with a total exposure of 200 minutes
The results, occurring during the Mkn 501 campaign, is also 
shown in Figure 3. The correlation between these measurements and the data 
collected immediately afterwards, during normal dark conditions, 
are excellent. We are therefore certain to add this twilight time as a further 
potential observing slot in case of urgent measurements, e.g. GRBs.  

\section{Prerequisites for operation at high background light levels}

The following precautions to minimise the impact of moonlight/twilight have 
to be
taken to prevent damage to the PMTs or producing unreliable data: 
(i) 
The PMTs should be operated with a gain of a few times  $10^4$, followed by
low noise AC-coupled preamplifiers. For this purpose,\ 6 - 8
stage PMTs would be ideal. Even
under severe background light the anode currents will be far below critical
values of fatigue and fast-ageing. A byproduct of low gain is a lower
number of positively charged ions liberated from the last stages. These
ions might eventually be accelerated and hit the cathode, thus creating
large pulses and in turn accidental triggers (Mirzoyan and Lorenz, 1996).
(ii) For prolonged moonshine operation, care should be taken to use good 
vacuum ($< 10^{-6}$ Torr) PMTs. This will prevent permanent damage by ions 
to the photo-cathode  and first dynodes.  
(iii) Due to the high noise rate the ATR will increase. Besides gain
reduction one can minimise the coincidence overlap time or introduce higher
level fast trigger systems such as so-called next-neighbour triggers. This
has recently been introduced successfully in the HEGRA telescopes. In order to
minimise the noise contribution to the data (and image analysis) the gating
time for the pulse height recording system should be minimised. For the
above presented observations we used 30 ns gates for the signal recording
ADCs.
(iv) Suppression of nearby scattered light can be achieved by using optimised
light collectors (Winston cones) in front of the PMTs, so that only light
coming from the mirror area (plus a small safety margin) is collected. This 
is also a standard feature of the HEGRA telescopes.
(v) Care should be taken, at large moon angles ($>90^{\circ}$), that no direct 
illumination of the PMTs by the moon occurs.
(vi) In order to minimise scattered light from the telescope frame, it should
be painted matte black.

Observation in presence of moon light increases the PMT's anode
current. This will accelerate the ageing of the PMTs. The dominant effect
is that due to intense electron bombarding the gain of the last dynodes
will decrease. For the used PMTs with CuBe dynodes it was found that the gain
drops by a factor of about 2 for an integrated anode current of 5 Coul/10 mm$^2$
dynode area. Very similar values were found for the 8$''$ PMTs of the wide angle
Cherenkov matrix AIROBICC 
(Karle, 1995). These  PMTs integrate the NSB over about 1 sterad. The
dynode area of the 8$''$ PMTs is about a factor 10 larger than that of the 9083A
PMTs. It should be noted that the reduction factor
fluctuates considerably from PMT to PMT and is very likely different for
PMTs from different manufacturers. From the about 1 year operation under
different light levels it was found that the gain change and trigger rate
are strongly correlated and the HT-gain correlation can be used to
reestablish the gain after ageing in a predictable manner. 
As mentioned above for minimising ageing it is obvious to operate the PMTs at the
lowest possible anode currents, respectively gain.

For the current operation mode of the CT1 camera it can be predicted that
the PMTs would have a lifetime exceeding 15 years when operating for 
about 500 h/y at half illuminated moon and a source separation of at least 20$^\circ$.

Note that the above arguments apply to a lesser amount also for observations around the galactic centre from southern locations. The central area of our galaxy is at least 10 times brighter than the dark celestial regions outside the milky way.

\section{Conclusions}
A simple technique with a fast reaction time, which can be used with imaging
 ACT telescopes to
increase the average observation time of an object, was
successfully tested. The only adjustment is a uniform decrease of a few 
percent in the HV
of the PMTs (0 to 13\% in this case). This can be realized within
seconds and will not affect the normal aging of the PMTs. 
Any telescope which operates at a moderate high PMT 
gain of a few times $10^4$ may use this technique.
Increases of the NSB up to a factor of 50 can be handled.
The technique
increases the threshold energy of the telescope by up to a factor of 2.2, 
depending on the HV reduction. 
It will 
allow more effective use of observation time (e.g. the early nights
during the waxing moon). Further investigations and refinements are
under discussion and it is believed that this technique could be
improved by optimising the supercuts for the various conditions discussed in 
this paper. 

\section*{Acknowledgements}
 This work is supported in part by the German Ministry for
Eduction and Research, BMBF, and the Spanish Research Organisation
CYCIT. The HEGRA Collaboration thanks also the Instituto d'Astrofisica de
Canarias for permission to use the site and continuing support. One of us,
C. Raubenheimer, thanks the Max Planck Society and the Oppenheimer Trust for 
financial support to
carry out this work. 

\section*{References}
Bradbury, S.M., et al. (1996), in Towards a Major Atmospheric Cerenkov Detector
-IV (Padova), 182\\
Chantell, M., et al. (1995), 24th Int. Cosmic Ray Conf. (Rome), {\bf 2}, 544\\
Dawson, B. and Smith, A. (1996), technical report GAP-96-034, University
of Adelaide\\
Karle, A,. et al. (1995), Astroparticle Physics 3, 321\\
Dobber, M.R. (1998), ESA preprint\\
Kranich, D., et al. (1997), Proc. 4th Compton Symp. (Williamsburg),{\bf 2}, 
1407 \\
Pare, E., et al. (1991), 22nd Int. Cosmic Ray Conf. (Dublin), {\bf 1}, 492\\
Petry, D., et al. (1996), Astron. \& Astrophys., {\bf 311}, L13\\
Petry, D., et al. (1997), private communication\\
Protheroe, R.J. et al. (1998), 25th Int. Cosmic Ray Conf. (Durban), {\bf 9}, in press\\
Quinn, J., et al. (1996), in Towards a Major Atmospheric Cerenkov Detector-IV
 (Padova), 341\\
Mirzoyan, R., et al. (1994), NIM A, {\bf 351}, 51\\
Mirzoyan, R., and Lorenz, E. (1996), in Towards a Major Atmospheric Cerenkov
 Detector-IV (Padova), 209\\
Rauterberg, G., et al. (1995), 24th Int. Cosmic Ray Conf. (Rome), {\bf 3}, 412\\
Schaefer, B.E., (1998), Sky \& Telescope (May), p 57 (submitted to Publ. Astr. 
Soc. Pacific) \\
Snabre, P., Saka, A., Fabre, B., and Espignat, P. (1998), Astropart. Phys., 
{\bf 8}, 159 \\
\vskip 10 pt

\vfil
\eject

\vskip 10pt
{\bf Table 1: Moonshine Observations of the Crab}\\
{(December 1996 - February 1997)}
\begin{center}
\begin{tabular}{|l|c|c|c|} \hline
 & OFF-region & Dark Night \\
 & $20^\circ < ALPHA < 80^\circ$ & OFF-data \\ [0.7ex] \hline 
Average zenith angle ($^\circ$) &  15.43 &  15.43\\
Observation time (min)  & 703 &     703 \\
Raw number of events  &  21230 &  21230 \\
Events after all cuts except ALPHA  & 668 $\pm$  26& 668 $\pm$   26 \\
Events after all cuts (ALPHA $\leq 10^\circ$) & 123 $\pm$ 11 & 123 $\pm$ 11 \\
Expected background events & 68 $\pm$  9&  81 $\pm$ 9\\
Excess events    & 55 $\pm$ 14 & 42 $\pm$ 13\\
Significance of excess ($\sigma$ )  &    4.0  &  3.4 \\
Excess rate (h$^{-1}$) &  4.7 $\pm$  1.2&  3.6 $\pm$   1.2 \\ [0.7ex] \hline
\end{tabular}
\end{center}

\vskip 20pt

{{\bf Table 2: Moonshine/Twilight (MS) and Dark Moon (DM) conditions and 
 observations of Mkn 501} (Zenith $< 59^{\circ}$, March - September 1997)}
\begin{center}
\begin{tabular}{|c|c|c|c|c|c|c|c|c|c|}\hline
$I_{moon}$ &NSB($30^{\circ}$)& $\Delta(HV)$  & $E_{th}$ & $N_{on}$   & $N_{b}$   & Exposure &Signifi-& Q   \\
(\%)& (nLambert) &(\%)& (TeV)   & & & T (h) &cance ($\sigma$) &  \\[0.7ex] \hline  
     & 54&  0 (DM)     & 1.2      & 5720 & 1187 & 183.0 & 51.2 & 12.3   \\
$<20$& 94&  0 (MS$^1$) & 1.2      &  647 & 150  & 17.9  & 18.6 & 13.0   \\
     &   & -6 (DM)    & 1.7      & 1306 & 407  & 60.9  & 22.3 &  8.3   \\
20-70&840& -6 (MS)    & 1.7      & 2151 & 927  & 74.9  & 20.1 &  8.2   \\
     &   & -9 (MS)    & 2.4      &  427 & 247  & 25.9  & 9.1 &  5.9   \\
70-90&1800&  -9 (MS$^2$)& 2.4      &  123 &  81  & 11.7  &  3.4 &  4.0   \\ 
90-95&2900& -13 (MS)    & 3.4      &  128 &  79  & 17.0  & 4.0 &  3.5  \\[0.7ex]\hline 
\end{tabular}
\end{center} 

$^1$ = including 3.3 h twilight observations; \\
$^2$ = Crab Nebula observations \\
$I_{moon}$ = moon illumination; \\
NSB ($30^{\circ}$) = visual NSB $30^{\circ}$ from moon;\\
$\Delta$(HV) = PMT voltage reduction\\
$E_{th}$ = threshold energy as calculated from PMT gain characteristics\\ 
$N_{on}$ = total events with ALPHA $<10^{\circ}$ \\
$N_{b}$ = expected background events with ALPHA $<10^{\circ}$ (see text)\\

\vfil
\eject
\begin{figure}[thb]
\centering
\epsfig{file=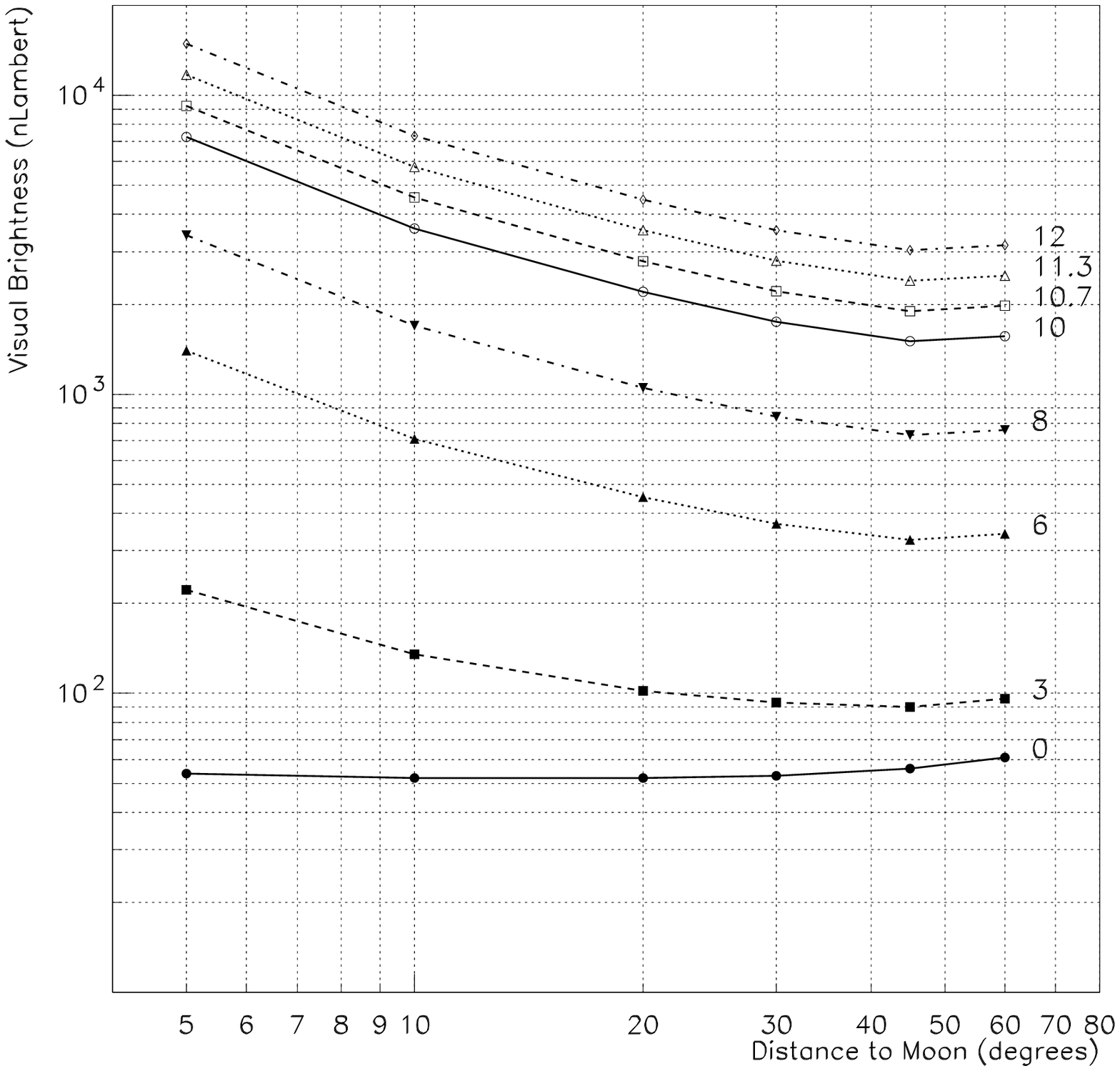,width=15cm}
\end{figure}

{\it Figure 1: The visual (300 - 900 nm) Night Sky Brightness as a function 
of angular separation between the moon and the observed object (moon angle),
 and the phase 
of the moon (expressed in days after new moon). The calculations used the 
model of Schaefer (1998).}
\vfil
\eject
\begin{figure}[thb]
\centering
\epsfig{file=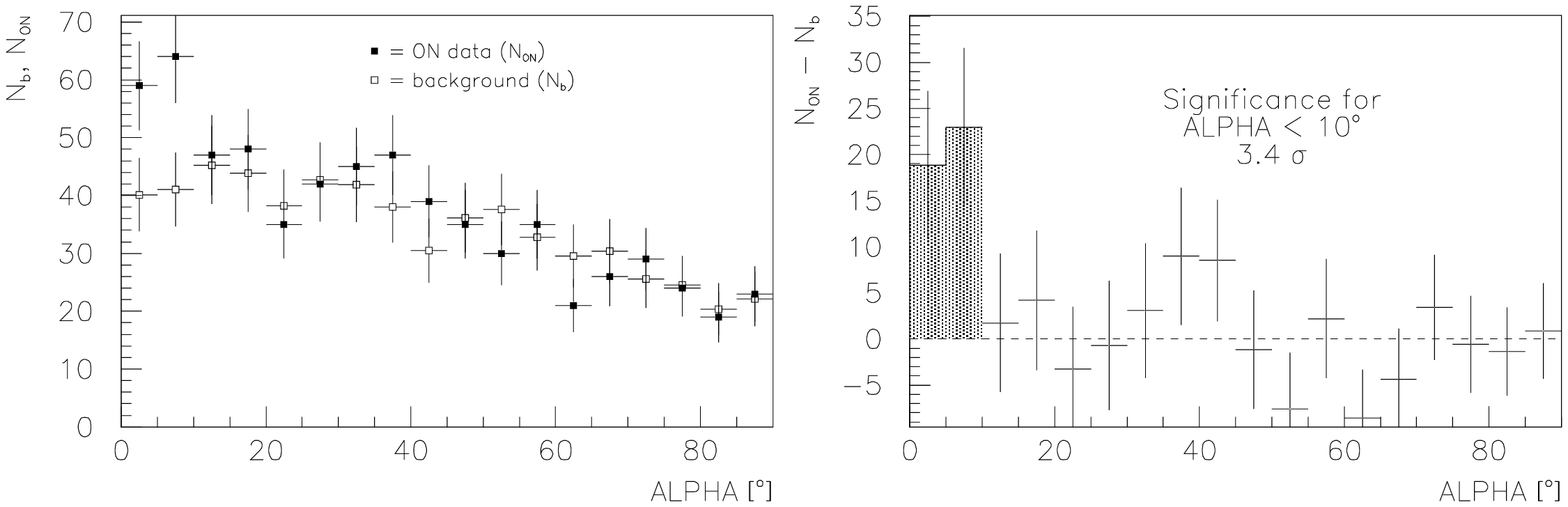,width=6cm,angle=-90}
\end{figure}

{\it Figure 2: The ALPHA-distribution for the 11.7 hours of moonshine 
observations of
the Crab Nebula. The data were obtained with the HEGRA CT1 telescope
at zenith angles smaller than 30$^\circ$. The filled symbols present the Crab 
measurements and the open symbols the background as obtained from previous
observations. An ALPHA-cut at 10$^\circ$ 
resulted in a significance
of 3.4 $\sigma$ at a threshold of $\sim$2.4 TeV.}
\vfil
\eject

\begin{figure}[thb]
\centering
\epsfig{file=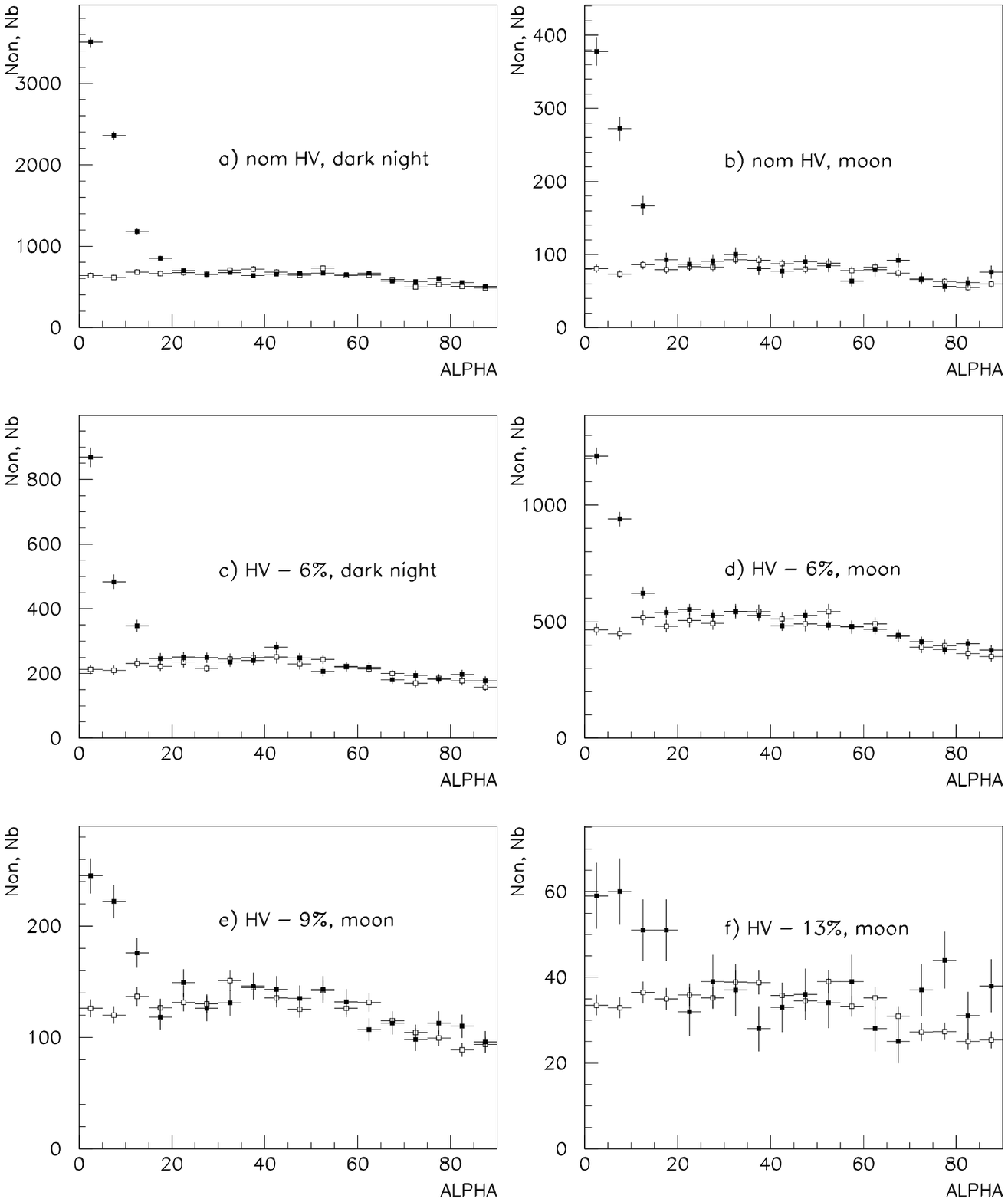,width=13cm}
\end{figure}

{\it Figure 3: ALPHA-distributions for the various categories (as indicated) of Mkn 501 
observations during 1997 with the HEGRA CT1 telescope. Full symbols represent 
the actual ON-source measurements and the open symbols are normalised 
OFF-source observations from Petry et al.(1997). The normal supercuts 
analysis was applied to the raw data.  
A flattening of the ALPHA-excess, with 
increasing HV reduction, is evident.}
\vfil
\eject

\begin{figure}[thb]
\centering
\epsfig{file=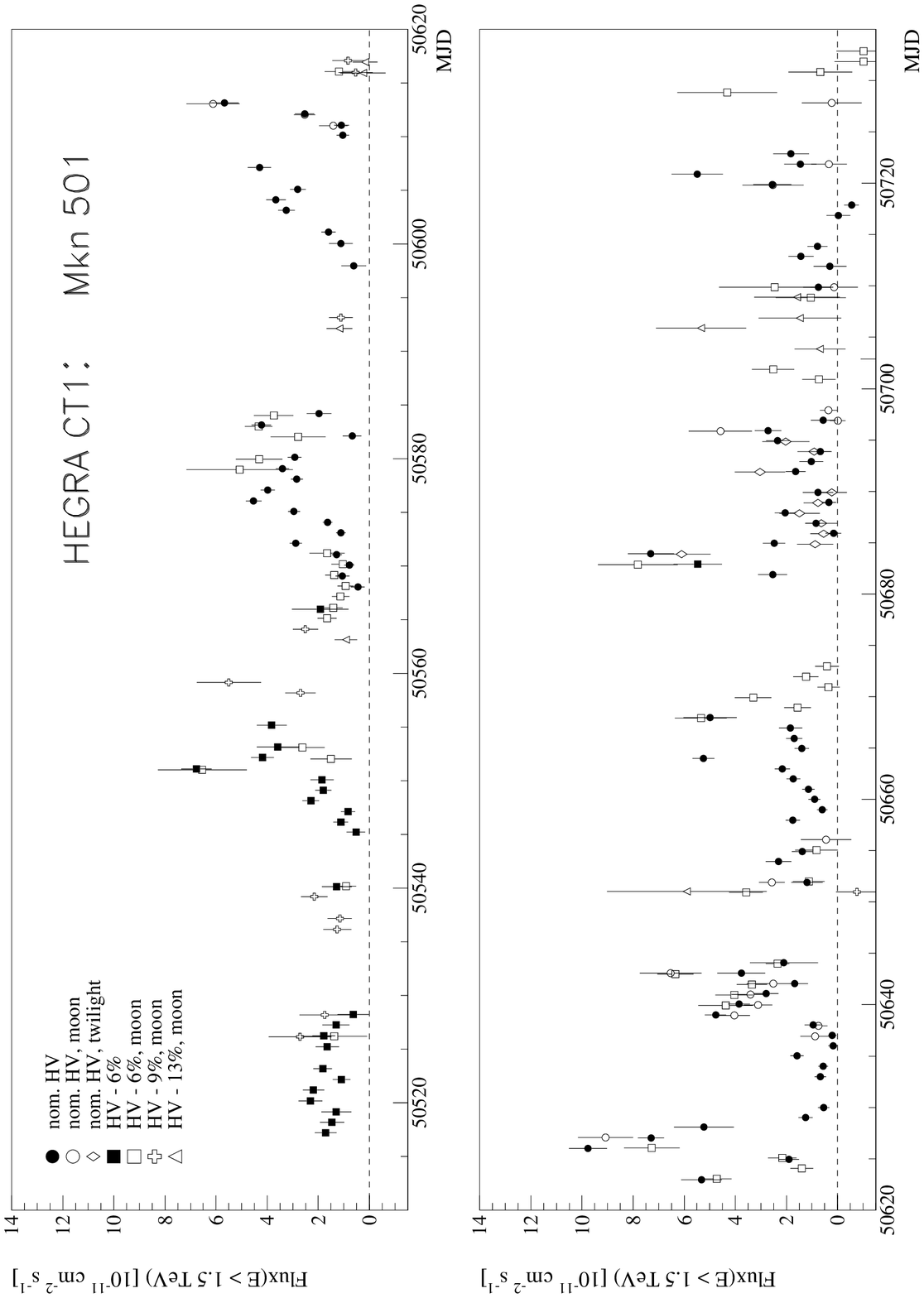,width=12cm}
\end{figure}

{\it Figure 4: The 1997 light curve of Mkn 501 (preliminary), as observed by 
the HEGRA 
CT1 telescope 
at an energy threshold of 1.5 TeV.
Open symbols represent moonshine/twilight observations whereas full symbols 
represent dark moon measurements.
For the flux extrapolation, a power law coefficient of $-1.5$ has been used.}
\vfil
\eject

\begin{figure}[thb]
\centering
\epsfig{file=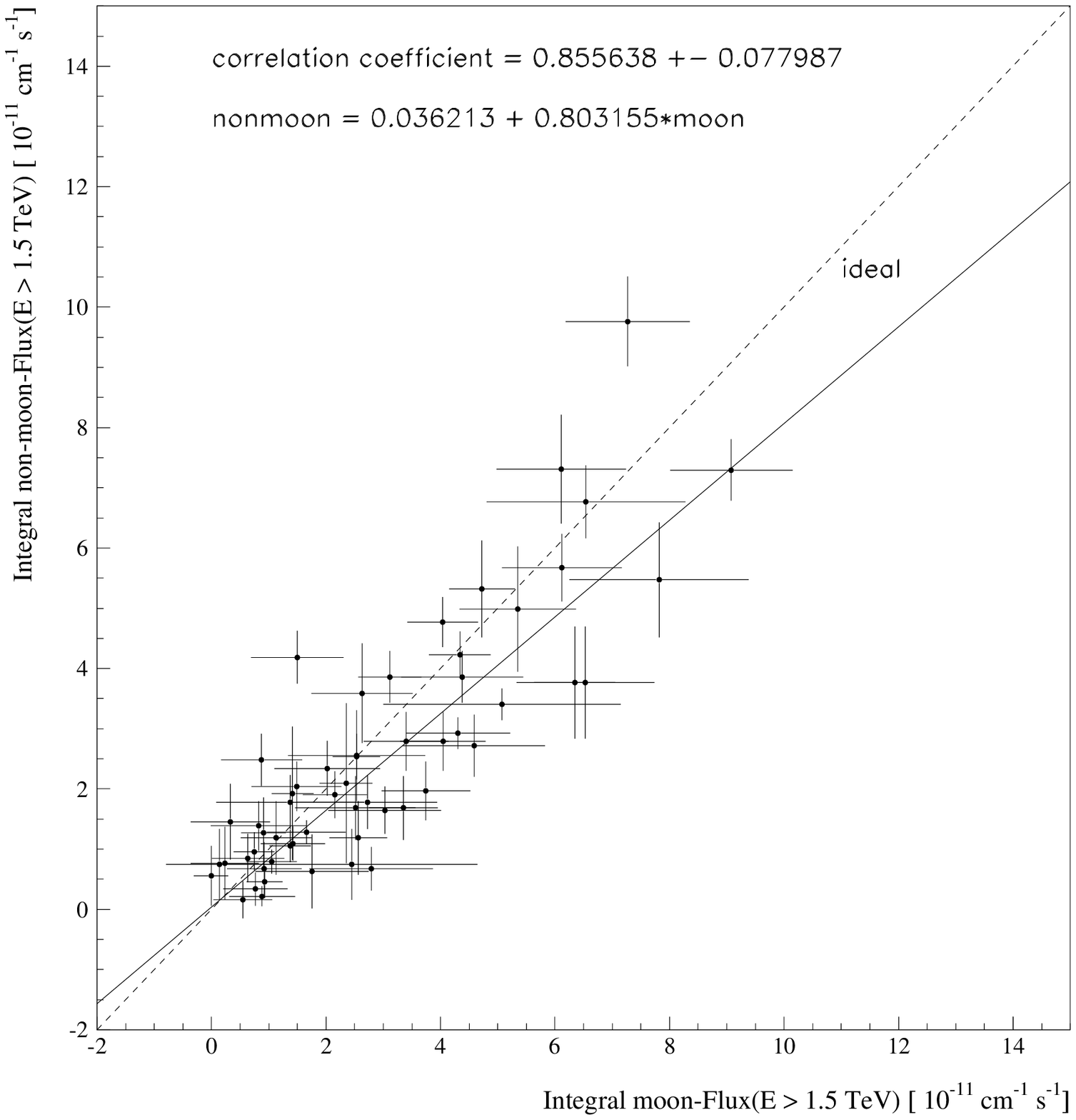,width=14cm,angle=-90}
\end{figure}

{\it Figure 5: The correlation between dark moon and moonshine/twilight fluxes
at 1.5 TeV for the 
57 pairs of measurements on Mkn 501, which occurred during the same night.}

\vskip 10pt

(moon1.tex)

\end{document}